\documentclass{elsart}

\usepackage{amsmath,graphicx}

\setkeys{Gin}{width=0.7\textwidth} % Size figs to this width by default.

\journal{Physica B}

\begin{document}
    % Definition of title page:
    \begin{frontmatter}
	\title{Analytic Conditions for Targeted Energy Transfer between 
    Nonlinear Oscillators or Discrete Breathers}
\author[Saclay]{S. Aubry}, \author[Saclay,Heraklion]{G. Kopidakis},
\author[Saclay]{A.M. Morgante}, \author[Heraklion,forth]{G.P. Tsironis}
\address[Saclay]{Laboratoire L\'eon Brillouin (CEA-CNRS), CEA Saclay\\
91191-Gif-sur-Yvette Cedex, France}
\address[Heraklion]{Department of Physics, University of Crete\\
P.O. Box 2208, 71003 Heraklion, Crete, Greece}
\address[forth]{Foundation for Research and Technology-Hellas\\
P.O. Box 1527, 71110 Heraklion, Crete, Greece}

\begin{abstract}

It is well known that any amount of energy injected in a harmonic oscillator
which is resonant and weakly coupled with a second harmonic oscillator,
tunnels back and forth between these two oscillators. When the two
oscillators are anharmonic, the amplitude dependence of their frequencies
breaks, in general, any eventual initial resonance so that no substantial
energy transfer occurs unless, exceptionally, an almost perfect
resonance persists. This paper considers this
interesting situation more generally 
between two discrete breathers belonging to two weakly 
coupled nonlinear systems, finite or infinite.
A specific amount of energy injected as a discrete breather 
in a nonlinear system (donor)  which is weakly coupled to another
nonlinear system (acceptor) sustaining another discrete breather,
might be totally transferred and oscillate back and forth between
these donor and acceptor breathers. The condition is that a
certain well defined detuning function is bounded from above and below 
by two coupling functions. This targeted energy transfer is selective,
i.e., it only occurs for an initial energy close to a specific 
value.
The explicit calculation of these functions in complex models with
numerical techniques developed earlier for discrete breathers,
allows one to detect the existence of possible targeted energy
transfer, between which breathers, and at which energy.
It should also help for designing models having desired
targeted energy transfer properties at will.
We also show how extra linear resonances could make the energy
transfer incomplete and \textit{irreversible}. Future developments
of the theory will be able to describe more spectacular effects,
such as targeted energy transfer cascades and avalanches, and 
energy funnels.  
Besides rather short term applications for artificially built devices,
this theory might provide an essential clue for understanding 
puzzling problems 
of energy  kinetics in real materials, chemistry, and bioenergetics.
\end{abstract}
\end{frontmatter}

\section{Introduction: a brief review of discrete breather theory}

It is a great pleasure to contribute to the volume dedicated
to Professor E. N. Economou with work that attempts to bridge
the gap between localization and propagation in complex discrete
systems. In the 
long-standing contributions of Prof. Economou, localization
is induced via quenched disorder while propagation is a result
of either higher dimensions or nonlinearity.  In our present work,
coherent energy transfer occurs as a result of judicious interlocking
of disorder with nonlinearity and an appropriate exploitation of 
the concept of nonlinear resonance.
Resonance is the basic principle for energy propagation in spatially 
periodic linear systems. The simplest example in textbooks is obtained for 
two coupled identical harmonic oscillators. In this case, it is well known 
that any amount of energy deposited on
one of the two oscillators tunnels back and forth between these two oscillators
with a frequency proportional to the coupling.
On the other side, in order to keep the energy localized in a given harmonic oscillator,
the resonance between this oscillator and the other harmonic 
oscillators must be broken. This situation can be reached in a
harmonic infinite system for an impurity mode or, more generally,
by breaking translational invariance, for instance through strong enough disorder
or by quasiperiodic modulations. 
Roughly speaking, finding a harmonic oscillator that is well resonant 
within the frequency interval $\delta \omega$ with another oscillator
generally requires to go far from the initial one, typically at a 
distance $\delta \omega^{-1/D}$ in a model at dimension $D$. But since the effective
coupling between these two almost resonant oscillators drops exponentially as a 
function of their distance, the resonance is generally not sufficient for allowing
a substantial energy transfer. This effect blocks energy propagation
and causes the Anderson localization of the eigenmodes
\cite{And58,Economou}.

When the system becomes nonlinear, this picture is drastically changed
 because the frequency of an anharmonic oscillator depends on its amplitude.
 Concerning the time-periodic solutions, we have shown that nonlinearity plays a
``double game" \cite{KA00,KAI99,KAII00}. First, by adjusting appropriately the 
amplitude of the 
nonlinear oscillators, nonlinearity could restore resonances between anharmonic 
oscillators which were lost in the linearized system. Indeed, small 
nonlinearity is sufficient to generate exact extended solutions 
in random systems while the whole linear spectrum is localized. 
Although these solutions appear spatially intermittent, they can propagate some 
nonvanishing amount of energy (much smaller though than
standard uniform plane waves).
Their existence demonstrates that random systems with a small 
nonlinearity can propagate energy while the same systems without 
nonlinearity cannot.
Second, with different amplitudes for the nonlinear oscillators,
nonlinearity can also maintain (or magnify) the lack of resonance
of a localized vibration with the  other modes. Thus, the  same system
can also sustain exact localized nonlinear modes. 
Moreover, the existence of these modes may not require randomness.
Mode localization can also be achieved in a nonlinear spatially periodic system provided
the frequency and its harmonics are nonresonant with the extended linear modes
(i.e., they do not belong to the linear phonon band).
These modes are called discrete breathers  \cite{ST88,CP90}.

Discrete Breathers (DBs), also called Intrinsic Localized Modes, are 
spatially localized time-periodic solutions
of discrete  classical nonlinear dynamical systems 
 which could be finite or infinite in arbitrary dimension,
spatially periodic or random, etc.
Many early works predated their discovery.
For example, polarons are nonlinear objects belonging to the family of DBs 
introduced long ago by Landau \cite{Lan33}.
The concept of \textit{local modes} was independently introduced in chemistry also
long ago \cite{BS26,E29,OE82} for small molecules.
In that case, discreteness was implicitly necessary.
Incidentally, these solutions are
recognized to be  important for understanding the quantization of 
molecular vibrations \cite{Far96}.  
 
Renewed interest on these nonlinear concepts but within the framework of 
integrable models which have exceptional (but sometimes
physically misleading) properties started during the 1970's.
The self-trapping equation introduced later in the mid-1980's
already revealed some features characteristic of DBs but
only in special cases (see a recent interesting review by A. Scott \cite{Sco99}).
Many interesting physical ideas of applications in 
physics were proposed, but were limited by flaws
of the underlying theories. They could be reconsidered and
amended using the more universal concept of DB.

It was only in 1988 that Sievers and Takeno \cite{ST88}
claimed the existence of DBs as long lifetime solutions in general models,
finite or \textit{infinite}, on the basis of approximate analytical and numerical
calculations (see review \cite{FW98}). Actually, their claims were definitely
confirmed later by a 
rigorous proof of their existence as exact solutions  in Klein-Gordon 
systems under rather general conditions of non 
resonance \cite{MA94}. Moreover, DBs were proven to be linearly stable
and robust to model perturbations (i.e., not restricted to specific 
models). Subsequently, the scope of validity of this proof  has been extended
to more general models with optical or acoustic phonons and with 
arbitrary complexity in any dimension 
\cite{MS95,Aub97,SM97,LSM97,Aub98}. This general theory also brought
new methods for systematic and accurate numerical investigations 
\cite{MA96}.

The  existence of DBs essentially requires \textit{discreteness and nonlinearity}
(randomness is not necessary despite the fact that DBs could persist in its 
presence). Thus, these conditions fully take into account 
two essential characteristics of real matter. It is discrete, because
it is made of atoms, and nonlinear, because the interatomic potentials are 
nonlinear.  DBs may exist generically as exact solutions 
when the frequency of a local mode and \textit{all of its harmonic} 
frequencies either  belong to phonon gaps or are above 
the whole phonon spectrum. Otherwise, these local modes would radiate away
their energy by the fundamental frequency or some harmonics and 
thus survive only over a finite time. However, if the radiation occurs by high order 
harmonics, the lifetime of this local mode at 0 K may be long but 
nevertheless finite.  
Thus, it is clear that DB existence as \textit{exact 
solution} crucially depends on the gap structure of the phonon spectrum.
An exact DB solution at a given frequency (or action) may exist in one model 
but may disappear in another very similar model just because
some high order harmonic becomes resonant with phonons.
Another essential property is that  DBs come as one
parameter family which can be parametrized by their 
frequency (or, better, their action which can be used for semiclassical 
quantization)
\footnote{The DBs we discuss here are extraband DBs,
to be distinguished from intraband DBs defined in Refs.\cite{KA00,KAI99,KAII00}.
Intraband DBs may only exist when the phonon modes are 
spatially localized in quasicontinuation of the Anderson modes.
The essential difference is that the frequencies (or harmonics)  of
intraband DBs lie inside the phonon spectrum but then these frequencies are 
restricted to fat Cantor sets. As a consequence, the intraband DBs do not form continuous 
families versus frequency.}.

DBs may have a lot of other amazing properties not fully explored yet but
which depend on the  considered models. They scatter elastically  
linear phonons (without energy exchange) like static impurities.
They may filter phonon noise coloring the spectrum \cite{CAF98}.
Inelastic interactions with small amplitude phonons (generally) occur at 
higher order. Then, DBs may slowly decay while in other situations 
unexpectedly they may grow \cite{JA00} by pumping energy from the background. 
They are generally pinned to the lattice but
may become highly mobile in some special situations \cite{CAT96} and then
they contribute to energy propagation instead of trapping it.
They can also react by fission or fusion \cite{AC98}.

Because they are \textit{exact solutions} at 0 K, and
the interaction between DBs drops very fast as a function 
of their distance, at low temperature the only physical way 
for a quasi-isolated DB to decay 
is by the small interaction with its \textit{own phonon background}.
Actually, the real spectrum of this phonon background is essential
for determining the true DB lifetime. Phenomenological interactions
modeled for example by a \textit{white} Langevin noise and damping 
(or, equivalently, a coupling with a large collection of harmonic oscillators 
with a uniform structureless spectrum), systematically destroy the DB as an exact 
solution simply because the phonon spectrum extends to 
infinity without gaps. Thus, the lifetime obtained by such approaches  
widely used in physics, is nothing but the result of the 
assumptions that are made and might be physically misleading. 

Consequently, when the temperature is sufficiently low (compared to the DB energy)
and the DBs sufficiently far apart, energy trapping 
by DBs may be substantial and persist over extremely long 
time in apparent violation of the standard Fourier law. 
Therefore, the most important feature characterizing (pinned) DBs 
(and distinguishing them from ordinary phonons and local modes), is
that in appropriate 
conditions where they can be created  (by external excitations or
by other processes breaking initially the thermal equilibrium), they may 
spontaneously appear  and \textit{persist} as spatially 
localized  dynamical structures with unusually \textit{long lifetime}.

Numerical simulations clearly demonstrate that in systems
where linearly stable DBs exist, they show up spontaneously under
thermal shocks \cite{TA96,BVAT99} or when starting from initial states 
far from thermodynamic equilibrium \cite{RABT00}.  
They trap energy over very long time and  generate unusual energy relaxation
(stretched exponentials). 
This long lifetime of DBs should be relevant for understanding many
anomalous energy relaxations  especially in disordered matter 
(glasses, polymers, biopolymers)
which up to day do not have yet truly satisfactory interpretations.

Up to now, experimentalists did not pay much attention to this new 
paradigm in physics. One of the reasons might have been that DBs 
could be easily confused with early concepts
inherited from the theory of integrable systems such as (nondiscrete)
breathers and  solitons \footnote{They only exist in
very special integrable models 
and, moreover, their properties are fragile to perturbations in contrast
with the highly 
robust DBs.}(perhaps because of the terminology). Otherwise, the
localization or the anomalous relaxation of energy  
observed in real experiments, are usually interpreted by
randomness, metastable states etc. 

In spite of the general belief, some experimentalists recently observed 
DBs with different techniques and in different systems,
for example in coupled nonlinear optical wave guides
\cite{ESMBA98,MPAES99}, in coupled arrays of Josephson junctions 
\cite{TMO00,BAUFZ00}, in some nonlinear materials
\cite{SBLSS99}, in magnetic systems \cite{SES99} and even possibly in myoglobin 
\cite{XMA00}.

After this brief review we now come to the main aim of this paper which is to 
present a novel 
property termed targeted DB energy transfer that enables
possible perfect transfer of energy between two DBs.
This idea was already briefly suggested in \cite{Aub98b} and considered in term 
of phase dynamics in \cite{AMS00}. We give here a precise analytic treatment
of this effect through a generalized dimer model. We should point out that our
results do not agree with the scenario imagined in \cite{AMS00}.
Although, this phenomenon can only exist in appropriate  models
and for appropriate DBs, it could become the key for understanding 
puzzling phenomena of well-focused energy transfer in physics and
especially in  biomolecules.

\section{Generalized Dimer Model for the Energy Transfer between two Discrete 
Breathers}

The energy tunneling between two resonant harmonic oscillators is 
described through the linear superposition of the symmetric and 
antisymmetric modes. This is of course not possible when the oscillators are 
(strongly) anharmonic. We show in this section that we can model the 
energy transfer in this case by an integrable generalized dimer model. We 
shall study  more 
generally the energy transfer between two DBs so that the energy
transfer between two oscillators becomes a particular case.

We consider an arbitrary anharmonic system (D) (finite or infinite)
called \textit{donor  molecule} described by some set of
variables and conjugate variables $\{p_{i}^{D},q_{i}^{D}\}$ and
a Hamiltonian $H_{D}(\{p_{i}^{D},q_{i}^{D}\})$. We assume that 
this molecule can sustain DBs and select one of them
called donor DB.

As we know, DBs come as continuous family
parametrized for example through their action (which is
$I_{D}=\int \sum_{i} p_{i} d q_{i}/2\pi$ integrated on the loop 
representing the DB in the phase space). Then, DB's energy 
$H_{D}(I_{D})$ becomes a function of action and 
its frequency is $\omega_{b}= dH_{D}/dI_{D}$.
We also assume that this DB family is gapless, i.e.,
$H_{D}(I_{D})\rightarrow 0$  for $I_{D} \rightarrow 0$
when the DB is vanishing at zero energy\footnote{Note that DBs in
space-periodic systems
in more than one dimension cannot be gapless \cite{FKM97}
but we essentially consider here systems where the space periodicity
is broken by the presence of the donor or acceptor sites.}.

We consider a second arbitrary anharmonic system (A) (finite or infinite)
called \textit{acceptor  molecule} with properties similar to the
donor molecule. We select on the acceptor molecule a family of acceptor 
DB with action $I_{A}$ and energy $H_{A}(I_{A})$. As with the donor system,
we assume that acceptor DB is also gapless.

Subsequently, we introduce a weak coupling between the donor molecule
and the acceptor molecule. 
These two systems could physically represent two independent molecules 
or two linked parts of the same macromolecule relatively far from one 
another. The weak coupling interaction could either come from contact
(hydrogen bonds, van der Waals forces), from (screened) Coulomb 
interactions, could be mediated by the solvent, or come
directly from the intermediate link of the macromolecule if any, etc.

We would like to find the conditions for complete transfer of the energy of 
a DB from donor molecule with action $I_{T}$, after some time, to a DB on the
acceptor molecule.
Since the coupling is weak, this transfer will be slow.
Then, at any intermediate time $t$, the dynamics of the system will be well 
described over a time scale long  compared to the period
by two DBs with action $I_{D}$ on the donor
and $I_{A}$ on the acceptor.  This solution is well represented
in terms of loop dynamics \cite{AC98} as an almost invariant closed loop in the phase space
which projects like DBs on the isolated molecules in
the complementary donor and the acceptor subspace.
Thus, the action of this loop is nothing but the sum of the actions of 
these two
isolated DBs. Because of the weak coupling between donor and acceptor 
systems, this loop evolves slowly
in the phase space. Liouville theorem
states that the action of a loop evolving in the phase space
is time-constant. Thus, we have 
\begin{equation}
    I_{T}=I_{D}+I_{A}.
    \label{eq:consact}
\end{equation}
Another necessary condition is the conservation of the total energy 
$E_{T}=H_{D}(I_{D})+H_{A}(I_{A})$ yielding 
\begin{equation}
    H_{D}(I_{D})+H_{A}(I_{T}-I_{D}) = E_{T}
    \label{eq:consenerg}
\end{equation}
which is a function independent of $I_{D}$ for $0 \leq I_{D} \leq I_{T}$.
Differentiating (\ref{eq:consenerg}) with respect to $I_{D}$
readily yields that the frequencies $\omega_{D}=dH_{D}/dI_{D}$ of the donor
DB and  $\omega_{A}=dH_{A}/dI_{A}$ remain identical during the energy 
transfer, i.e., the two DBs remain resonant.

Actually, this condition (\ref{eq:consenerg}) does not need to be perfectly fulfilled
(since the coupling energy should also be involved in the energy 
conservation) but we shall assume that it is almost fulfilled.
We wish to describe the coupled donor-acceptor system through an effective 
Hamiltonian expressing the energy transfer only as a function of the variables 
$I_{D}$ and $I_{A}$ and their
conjugate variables $\theta_{D}$ and $\theta_{A}$. 

However, the restriction to two pairs of variables will be valid
only if condition (\ref{eq:consenerg}) is not fulfilled precisely or even approximately 
by any other DB of the acceptor molecule for the same total 
action $I_{T}$ and the same total energy $E_{T}$. In other words, we 
assume that the acceptor DB is unique. We shall assume also, but only for 
simplicity, that the donor DB is unique
\footnote{Otherwise we 
should have to involve other donor DBs in the return transfer.
This problem, which is also interesting and related to energy funneling, will be 
treated elsewhere.}. This condition is imposed for avoiding more complex 
situations with bifurcating transfer of energy to two or more 
acceptor DB. It is in principle fulfilled for molecules 
without any special symmetry.

A priori, the small coupling energy, $C(I_{D},I_{A},\theta_{D},\theta_{A})$ 
depends on  $I_{D} \equiv I_{0}+I$ and $I_{A} \equiv I_{0}-I$, on the angle difference
$\theta=\theta_{D}-\theta_{A}$ (conjugate to $I$) supposed to vary slowly because of the
above assumptions (weak coupling and almost resonance),
and on the total angle  $\theta_{0}=\theta_{D}+\theta_{A}$ 
(conjugate to $I_{0}$) which on 
the opposite varies fast at the scale of the DB periods.
It is then justified to average the coupling over the fast variable
$\theta_{0}$ (which varies practically linearly over relatively long time)
and drop the $\theta_{0}$ dependence.
Next, it is convenient to split the coupling energy as a sum
$C(I_{0},I,\theta)=C_{0}(I_{0},I)+C_{1}^{\prime}(I_{0},I,\theta)$,
where $C_{0}(I_{0},I)$ is the average of $C(I_{0},I,\theta)$ over $\theta$,
while the average of $C_{1}^{\prime}(I_{0},I,\theta)$ over $\theta$ is 
zero.

Then, the effective Hamiltonian regarding the two DBs of the coupled molecules
can be written in the form
\begin{equation}
    \mathcal{H}(I_{0},I,\theta)= H_{0}(I_{0},I) + V(I_{0},I,\theta)
    \label{eq:hameff}
\end{equation}
where 
$H_{0}(I_{0},I)=H_{D}(I_{0}+I)+H_{A}(I_{0}-I)+C_{0}(I_{0}+I,I_{0}-I)$
and $V(I_{0},I,\theta)=C_{1}^{\prime}(I_{0}+I,I_{0}-I,\theta)$
which has zero average with respect to $\theta$. Moreover,
for $I=\pm I_{0}$, either the acceptor DB or the donor DB 
vanishes which removes any $\theta$ dependence. Then,
$V(I_{0},\pm I_{0},\theta)=0$.

Since this Hamiltonian does not depend on $\theta_{0}$, 
$\dot{I}_{0}=-\partial \mathcal{H}/\partial \theta_{0}= 0$,
which implies that the total action $I_{D}+I_{A}=2I_{0}$ is conserved.
Because of the averaging over the fast variable $\theta_{0}$,
$\theta$ essentially represents the difference between the DB phases
on the donor and acceptor molecules.
Effective Hamiltonian (\ref{eq:hameff}) with $I_{0}$ a 
constant parameter is nothing but the Hamiltonian for the DB 
phase (see ref. \cite{AMS00} for a formal theory of phase dynamics).

For donor and acceptor molecules with explicitly given Hamiltonians
and coupling, it is possible to calculate numerically this effective
Hamiltonian and thus to predict the possibility of complete targeted 
transfer.
This Hamiltonian can be formally written as a generalized Discrete Nonlinear
Schr{\"o}dinger (DNLS) dimer model
when defining the complex variables $\psi_{D}=\sqrt{I_{0}+I} \quad 
e^{-i(\theta_{0}+\theta)/2}$ at the donor site $D$ and 
$\psi_{A}=\sqrt{I_{0}-I}\quad 
e^{-i(\theta_{0}-\theta)/2}$ at the acceptor site $A$. Then, the
conservation of the total action $I_{0}$ 
is equivalent to the conservation of the total norm 
$|\psi_{D}|^{2}+|\psi_{A}|^{2}$.
This approach could be easily modified when the frequency of one of the two 
DBs is resonant with a harmonic of the other one. We could then 
analyze more accurately for DBs the problem of Fermi resonance usually
considered for quasilinear modes.
This approach can be also extended when there are more than two weakly 
coupled nonlinear oscillators. We then obtain a generalized DNLS model
on a lattice. It is also extendable for quantum anharmonic oscillators.
We then obtain a quantum generalized DNLS model which is nothing but a boson model
on a lattice. In this case, the assumption of conservation of the classical action
becomes conservation of the number of bosons and
nonlinearity becomes many-body interactions. On this basis we can generalize
early calculations based on rotating wave approximations,
which were initially restricted to specific models (e.g. theory of the
Davydov soliton, see \cite{Sco99}). Expanding the obtained effective 
Hamiltonians at the lowest significant order simply yields standard quartic 
DNLS models. However, we also learn from our approach that these approximations
are only valid in the weak 
coupling limit, i.e., close to an anticontinuous limit, which was not
clearly stated in the early approaches. It is thus inconsistent to 
describe later these DNLS models from their continuous limit where 
they are integrable. These extended problems will be discussed later in further 
publications.

\section{Targeted energy transfer solutions in the generalized dimer model}

We search for solutions of the dynamical system defined by
Hamiltonian (\ref{eq:hameff}), which correspond to a total energy transfer.
These solutions are such that if at the initial time all the energy
is on the donor molecule, i.e., $I_{D}(0)=2 I_{0} \neq 0$ and $I_{A}(0)=0$, 
it will be completely transferred at a later time $t_{T}$ on 
the acceptor molecules, i.e., 
$I_{D}(t_{T})=0$ and $I_{A}(t_{T})=I_{D}(0)$. Then,
$I(0)=I_{0}$ and $I(t_{T})=-I_{0}$. 
We say by definition that when this situation occurs, we have targeted energy 
transfer.

Since for $I=I_{0}$, as well as for $I=-I_{0}$, the angle $\theta$ is 
undetermined, the topology of the phase space defined by
the conjugate variables $(I,\theta)$ is the one of a 2D sphere represented
in cylindrical coordinates
\footnote{The topology of this sphere was also considered in 
\cite{AMS00} for two weakly coupled pendula.}. Any point $\textbf{P}$
of this sphere can be represented by the two coordinates which are
the angular coordinate $\theta$ (longitude) of its projection in the 
$(x,y)$ plane and $I$ is its $z$ coordinate (latitude).
The poles of the sphere are the points where $I=\pm I_{0}$.

There are two time invariants for this Hamiltonian which are
$I_{0}$ and the total energy $E$. Then, 
\begin{equation}
     V(I_{0},I,\theta) + H_{0}(I_{0},I)= E
    \label{eq:orbit}
\end{equation}
defines contour lines  on the sphere $(I,\theta)$ which represent 
the orbits of the dynamical systems in the  2D phase space.

Since these contour lines are closed loops, $I(t)$ is time-periodic. 
Actually, the solution in the original system
consisting of the coupled donor and acceptor molecules, is 
a two site DB the period of which varies periodically
with the same period as $I(t)$. Thus, the global solution is 
quasiperiodic.

The dynamical solutions of the system can be classified in 
two types. Type 1 corresponds to the orbits on the sphere which are homotopic to 
zero, i.e., that can be continuously shrunk to zero without
crossing a pole. For these trajectories,
$\theta$ oscillates between two bounds. The limit case where $\theta$
does not oscillate, corresponds to exact DBs.
They correspond to extrema of the energy (\ref{eq:orbit}) on the 
sphere at fixed $I_{0}$.

Type 2 corresponds to the orbits which turn around the axis connecting 
the two poles. Then, $\theta$ rotates by $2\pi$ around the sphere
at each period of $I(t)$
\footnote{In the case of a nonresistive and noninductive Josephson
junction (JJ), which can be described by a dimer model, type I and class
II characterizes the two possible states of the JJ without DC current and 
with DC current, respectively \cite{FAM98}.}.

A trajectory corresponding to a targeted energy
transfer between the donor DB and the acceptor
DB is such that along the orbit, $I(t)$ 
varies between $I_{0}$ and $-I_{0}$. Thus, it is an orbit connecting 
directly the two poles of the sphere $(I,\theta)$.
It is straightforward to see that
a necessary and sufficient condition for having a
targeted transfer solution at action $I_{0}$, is that the system only
accepts type 1  solutions at this total action.

Let us explicit more carefully, this condition.
Since the two poles must be at the same energy, we have
\begin{equation}
    H_{0}(I_{0},I_{0})= H_{0}(I_{0},-I_{0})= E.
    \label{eq:nlrescond1}
\end{equation}
At the uncoupled limit $V(I_{0},I,\theta) \equiv 0$, this condition 
is fulfilled when (\ref{eq:consenerg}) is fulfilled. If 
$H_{0}(I_{0},I)$
is not constant, the contour lines are just circles at constant $I$
and are of type 2 and thus, we do not have targeted energy transfer.

When the coupling is turned
on, targeted  energy transfer requires that there exists
an action $I_{T}$ associated with an energy  $E_{T}$ 
such that Eq.(\ref{eq:nlrescond1})  is fulfilled for $I_{0}=I_{T}/2$
and $E=E_{T}$. 
Then, for $E=E_{T}$, this closed orbit 
on the sphere is defined by implicit equation
\begin{equation}
     V(I_{T}/2,I,\theta) = H_{0}(I_{T}/2,\pm I_{T}/2)- H_{0}(I_{T}/2,I)
     =\epsilon_{T}(I).
    \label{eq:orbit2}
\end{equation}
Function $\epsilon_{T}(I)$  vanishes at the poles for $I=\pm I_{T}/2$ and 
 characterizes the detuning between the donor and acceptor 
 DBs during the energy transfer. We term it 
 \textit{detuning energy function}. In some sense, $-\epsilon_{T}(I)$ 
extends the concept of Peierls-Nabarro energy barrier,
which concerns only static excitations, to  DBs, which are 
 dynamical  nontopological excitations. However, note now that the barrier
in energy might be positive or negative.
With no restrictions on the choice 
of the coupling functions $V(I_{T}/2,I,\theta)$,
it is rather easy to find sophisticated dependence on $\theta$
(but a bit unrealistic) , such that
there cannot exist any contour line connecting
the two poles.
Thus, it is convenient to require some properties for the $\theta$
dependence which also make the coupling physically reasonable
especially if this coupling is small.
We shall assume that it is sine-like and
that is it has only one maximum and one minimum per period
for $-I_{T}/2<I<I_{T}/2$.
This coupling function $V(I_{T}/2,I,\theta)$ has 
two (and only two)  zeros per period with respect to $\theta$ 
since its average is zero. 
More generally,  Eq.(\ref{eq:orbit2}) determines either two values modulo $2\pi$
for $\theta$ or no values. 
A necessary and sufficient condition for having two solutions
$\theta_{+}(I)$ and $\theta_{-}(I)$ which are continuous functions of 
$I$ on the sphere $(I,\theta)$ is simply
\begin{equation}
   V_{-}(I)=\min_{\theta} V(I_{T}/2,I,\theta) < \epsilon_{T}(I)< \max_{\theta} 
   V(I_{T}/2,I,\theta)=V_{+}(I)
    \label{eq:detuning}
\end{equation}

Functions $\theta_{+}(I)$ and $\theta_{-}(I)$ determine the two segments
of the closed contour line which connects the two
poles of the sphere  $(I,\theta)$.
The corresponding trajectory is a trajectory which realizes
a targeted energy transfer.
We note that during the time $t_{T}$, the phase difference between the donor 
and acceptor DBs $\theta_{\pm}(I)$
does not rotate over the trigonometric circle but varies in some 
interval.

We have just demonstrated (for a physically reasonable sine-like dependence 
of the coupling on the phase difference) that 
targeted energy transfer solutions do persist when the detuning energy
function is not strictly zero provided  the detuning energy function 
is bounded by the max and the min of 
the coupling function. When the detuning function is small, this 
coupling may be also small.

The targeted energy transfer solution at $I=I_{T}/2$
may be lost when varying the model parameters.
It suffices that condition 
(\ref{eq:detuning}) is no more fulfilled. At the threshold, there is a 
certain value $I_{b}$ of $I$ where $V(I_{T}/2,I,\theta)=V_{\pm}(I)$. 
Then, $\theta= \theta_{b}=\theta_{+}(I_{b})=\theta_{-}(I_{b})$.
Consequently, point $(I_{b},\theta_{b})$ is an extremum of 
the energy (\ref{eq:orbit}) on the sphere so that 
$I(t)=I_{b},\theta(t)=\theta_{b}$ is time-independent,
which in the initial system is time-periodic. They correspond to
exact DB solutions
of the coupled system made of the donor and acceptor molecules.
 
Thus, the threshold where targeted transfer disappears,
can be precisely associated with the appearance of a new DB
solution of the whole system. When approaching this threshold,
the targeted transfer solution becomes cnoidal, i.e.,
the time of transfer diverges because the energy transfer slows down 
in the vicinity of $I_{b}$ where the new DB solution will appear.

\section{Explicit solutions in the general quartic dimer model}

We now investigate explicitly the targeted energy transfer 
solutions in exactly solvable dimer models which describe quite well 
the general behavior in the weak coupling limit. 
DNLS Dimer models were studied in many early works \cite{KC87,KT87,GS98}
but only in special cases with no special attention to the detuning function.
Energy transfer solutions were already found but for relatively large coupling
in cases where the detuning function (\ref{eq:orbit2}) was not small. 
It is thus instructive to reconsider a more general quartic DNLS dimer model
as an illustration of our general theory. We consider a dimer with  Hamiltonian
\begin{equation}
        {\mathcal H} = \left(\frac{1}{2} \chi_{1}|\psi_1|^4 + 
        \omega_{1}|\psi_1|^2\right)
	+\left(\frac{1}{2} \chi_{2}|\psi_2|^4 + \omega_{2}|\psi_2|^2\right)
	 - \lambda (\psi_1 \psi_2^* +\psi_1^* \psi_2)
        \label{dimhamiltg}
\end{equation}
where $\chi_{1}$, $\chi_{2}$, $\omega_{1}$ and $\omega_{2}$ are 
parameters.  $\lambda$ is the inter-site coupling chosen positive for 
convenience. $(\psi_{1}^{\star},i\psi_{1})$ and 
$(\psi_{2}^{\star},i\psi_{2})$ are pairs of conjugate variables.
Defining first the new pair of conjugate
variables ($I_{1}$,$\theta_{1}$) and ($I_{2}$,$\theta_{2}$) by
$\psi_{1}=\sqrt{I_{1}} e^{-i \theta_{1}}$ and 
$\psi_{2} = \sqrt{I_{2}} e^{-i \theta_{2}}$
and next the pair of conjugate variables ($I_{0}=(I_{1}+I_{2})/2$,
$\theta_{0}=\theta_{1}+\theta_{2}$) and ($I=(I_{1}-I_{2})/2$,
$\theta=\theta_{1}-\theta_{2}$), an equivalent form of Hamiltonian 
(\ref{dimhamiltg}) is
\begin{equation}
  \mathcal{H}(I,\theta,I_{0},\theta_{0}) =  H_{0}(I_{0},I)+
   V(I_{0},I,\theta)
    \label{eq:energ}
\end{equation}
with 
\begin{eqnarray}
    H_{0}(I_{0},I) & = & \frac{1}{2} \chi_{0} 
  (I_{0}^{2}+I^{2}) + \chi I_{0} I + \omega_{0} I_{0} + \omega I
    \label{eq:dimint}  \\
    V(I_{0},I,\theta) & = & -2 \lambda \sqrt{I_{0}^{2}-I^{2}}\cos{\theta}
    \label{eq:dimcoup}
\end{eqnarray}
and the new model parameters 
\begin{equation}
    \chi_{0}= \chi_{1}+\chi_{2}, \quad
    \chi = \chi_{1}-\chi_{2}, \quad
    \omega_{0} =  \omega_{1}+\omega_{2}, \quad
    \omega = \omega_{1}-\omega_{2}
    \label{eq:omega}
\end{equation}

Complete targeted energy  transfer is obtained for
\begin{equation}
    I_{T} = -\frac{2\omega}{\chi}, \qquad  E_{T} = \frac{ \omega}{\chi} \left(
    \frac{\omega\chi_{0}}{\chi}-\omega_{0}\right)
    \label{eq:engtg}
\end{equation}
The detuning function (\ref{eq:orbit2}) and the coupling functions $V_{+}(I)$
and $V_{-}(I)$ in Eq.(\ref{eq:detuning}) are
\begin{equation}
    \epsilon_{T}(I)= \frac{1}{8}\chi_{0}(I_{T}^{2}-4 I^{2}) \quad \mbox{and} 
    \quad
    V_{+}(I)=-V_{-}(I)= \lambda \sqrt{I_{T}^{2}- 4 I^{2}}
    \label{eq:detunf}
\end{equation}
so that inequality (\ref{eq:detuning}) is fulfilled for
all $I$ when 
\begin{equation}
    \lambda >\left|\frac{\chi_{0} \omega}{4 \chi}\right|
    \label{eq:critcoup}
\end{equation}
Targeted energy transfer is obtained for any nonzero coupling
when $\chi_{0}=0$, i.e., when the nonlinear coefficients on the two
sites are opposite. On the contrary, it is never obtained when $\chi=0$, i.e.,
when the nonlinear coefficients on both sites are equal,
except when $\omega=0$. The case $\chi=0$, $\omega=0$, which 
was treated in Ref.\cite{KC87}, is special because $I_{T}$ is undetermined.  
Then, there are exact energy transfer solutions for any $I_{0}$ when 
the coupling is large enough, i.e., $\lambda> \chi_{0} I_{0}/4$. Exceptionally,
energy transfer is not selective in this case.

\begin{figure}[tbp]
    \centering
\includegraphics[width=0.4 \textwidth]{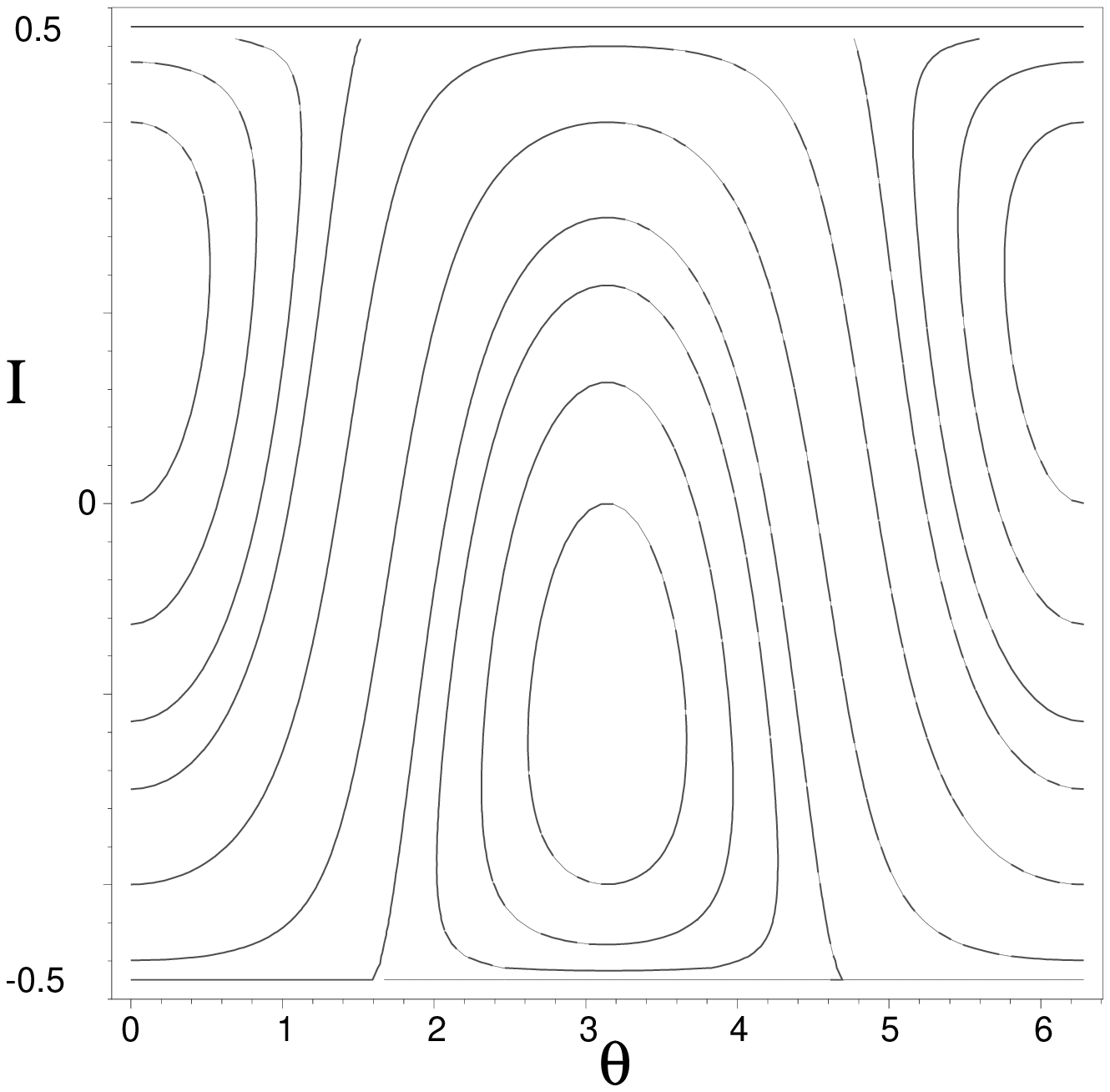}
\includegraphics[width=0.4 \textwidth]{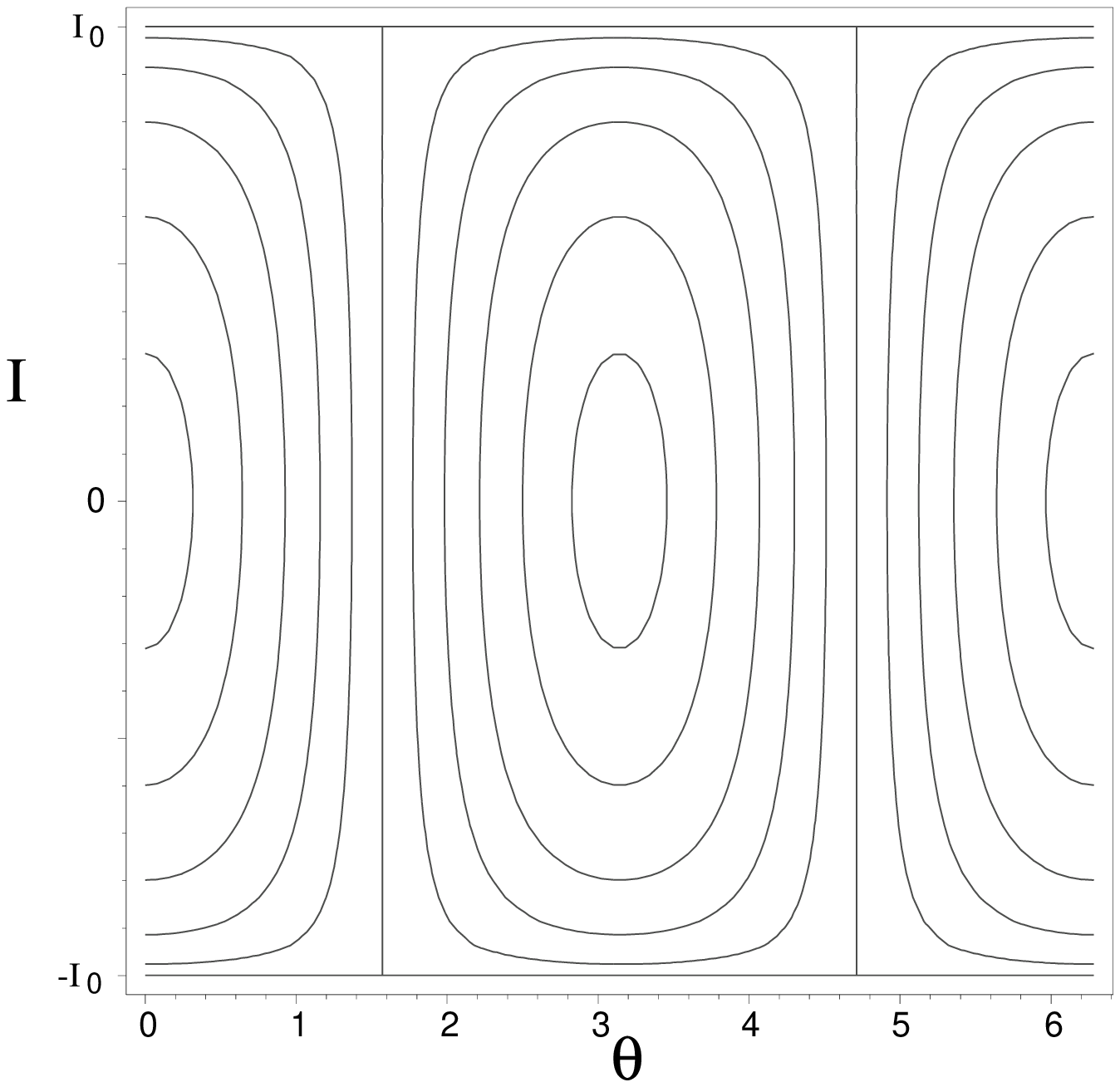}
 \includegraphics[width=0.4 \textwidth]{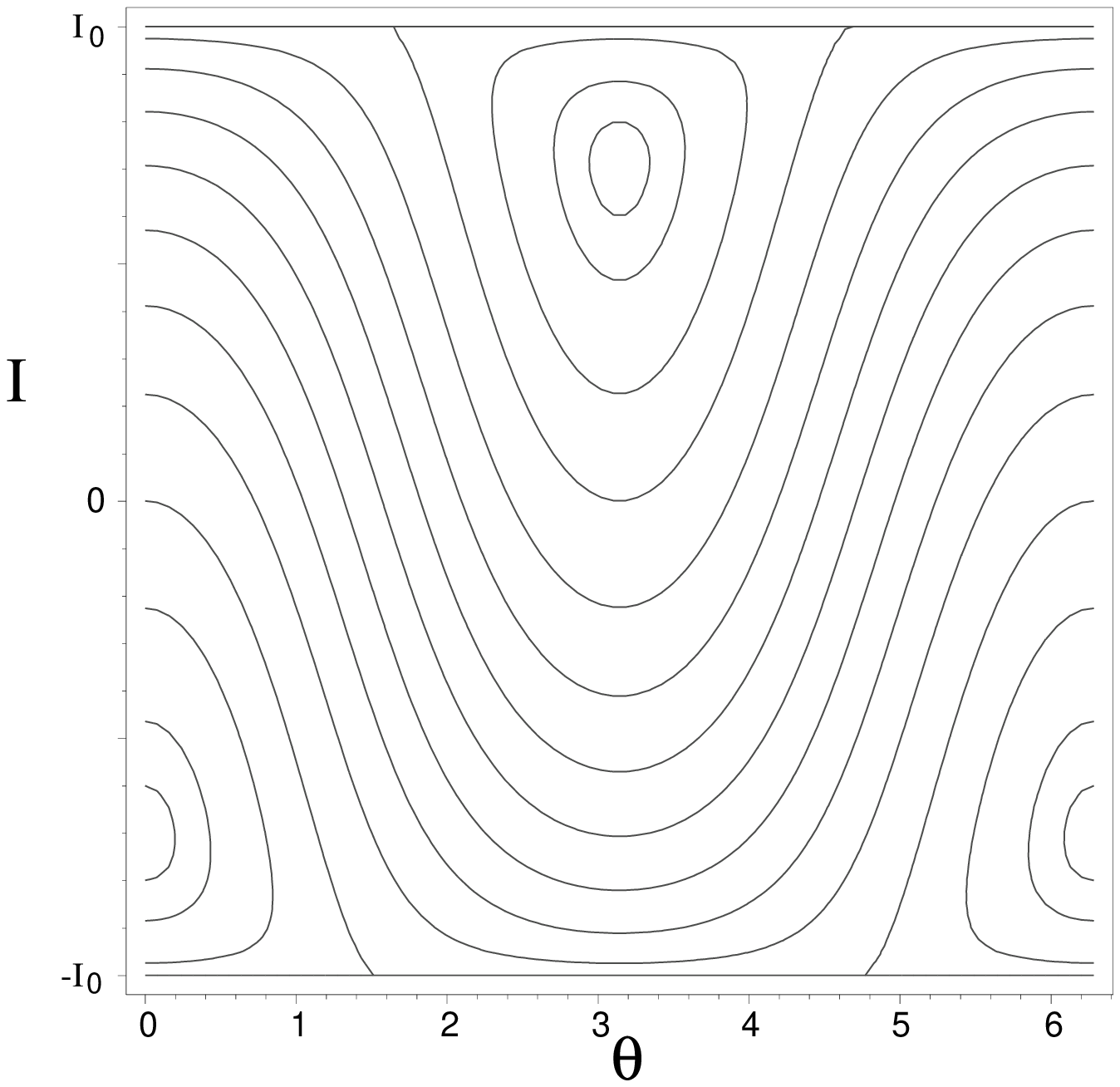}
    \caption{Cylindrical projections of the contour lines on the 
    sphere $(I,\theta)$ of Hamiltonian (\ref{eq:energ}) 
for $I_{0}=0.5$ (a), $I_{0}=1.0$ (b) (ideal targeted transfer)
, $I_{0}=2.0$ (c). Model parameters are
    $\chi_{0}=0$, $\chi=1$, $\omega_{0}=0$, $\omega=-1$,
    $\lambda=1/2$.}
    \label{fig1a}
\end{figure}

\begin{figure}[tbp]
\centering
\includegraphics[width=0.4 \textwidth]{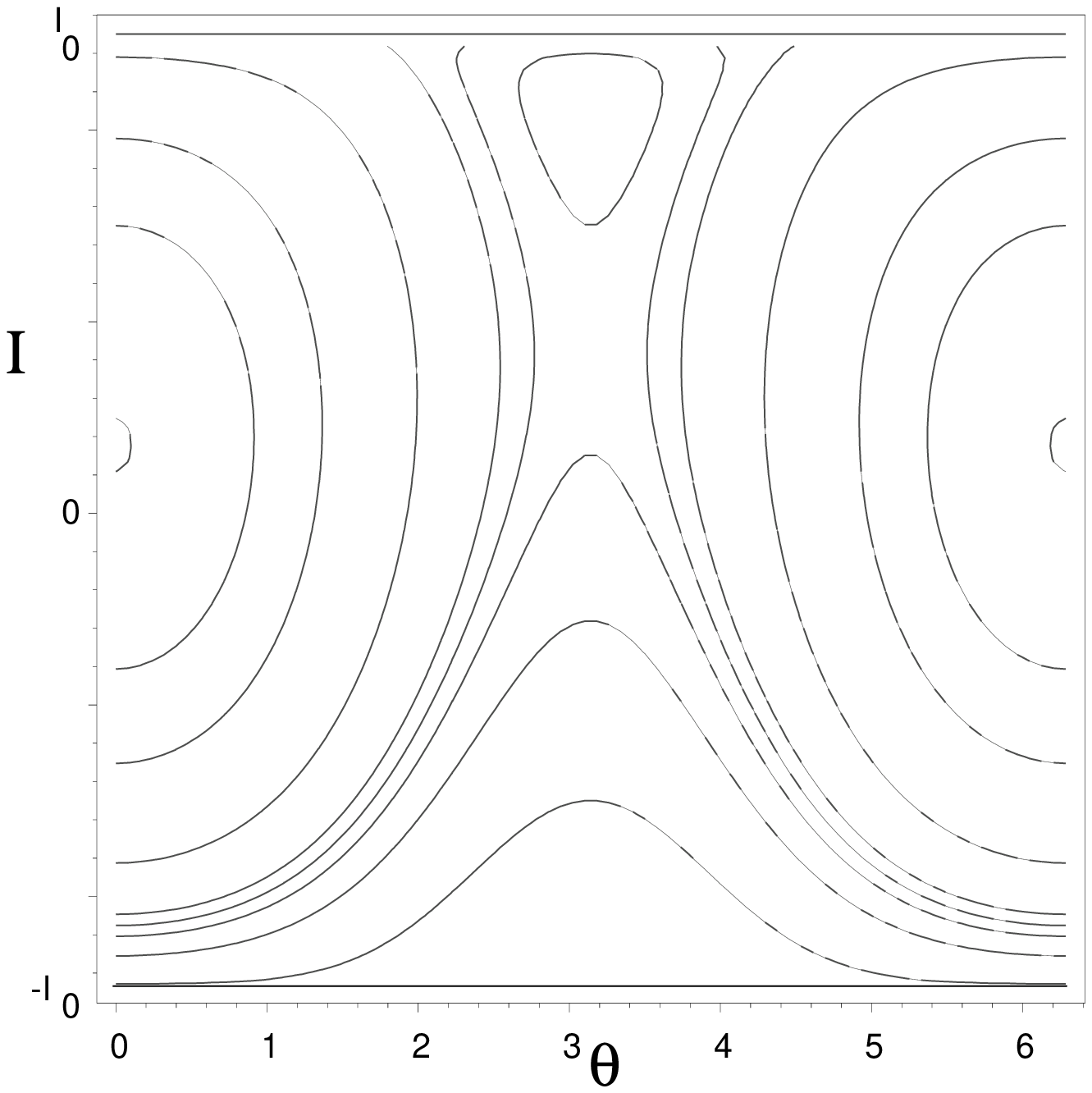}
\includegraphics[width=0.4 \textwidth]{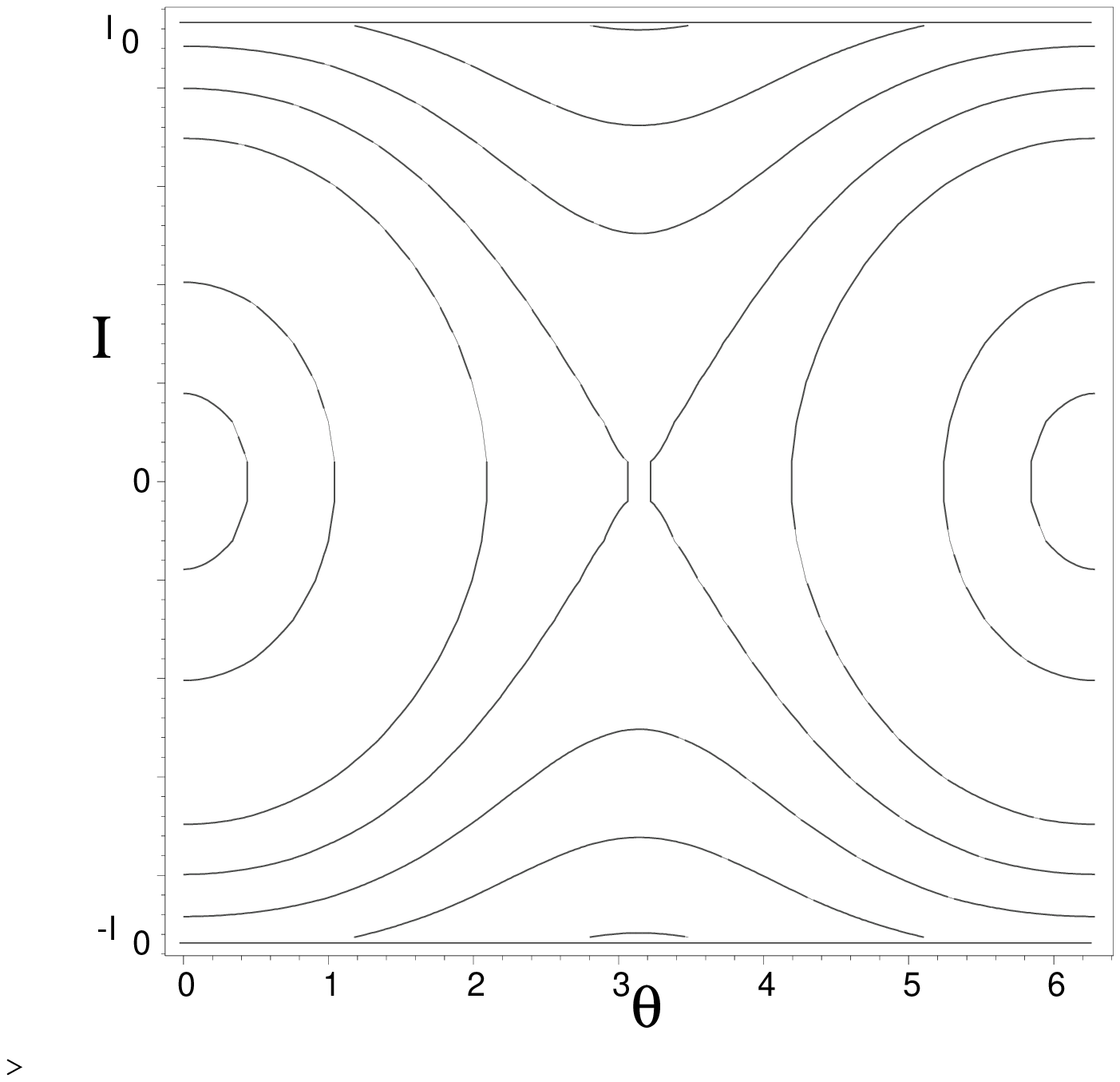}
\caption{Same as fig.\ref{fig1a} for $I_{0}=0.5$
(a) and $I_{0}=2.0$ (b) but with parameters
$\chi_0=5$, $\chi=1$, $\omega_0=0$
, $\omega=-1$, $\lambda=1/2$.}
\label{fig1b}
\end{figure}

Eq.(\ref{eq:orbit}) for Hamiltonian (\ref{eq:energ}) yields the model
trajectories which appear as contour lines on the sphere $(I,\theta)$ defined 
by 
\begin{equation}
    E= \frac{1}{2} \chi_{0} (I_{0}^{2}+I^{2}) + \chi I_{0} I 
    + \omega_{0} I_{0} + \omega I
    - 2 \lambda \sqrt{I_{0}^{2}-I^{2}}\cos{\theta} 
    \label{eq:costheta}
\end{equation}
They are shown for several values of the total action $I_{0}$ and for 
a choice of the model parameters such that targeted  energy transfer occurs
for a certain initial 
action $I_{0}$ and energy $E$ (cf. Fig.\ref{fig1a}).
An example is also shown when targeted energy transfer never occurs at any
action $I_{0}$ and energy $E$ (cf. Fig.\ref{fig1b} ).

Eliminating $\theta$ between Eq.(\ref{eq:costheta}) and 
the Hamilton equation
\begin{equation}
    \dot{I}=- \frac{\partial \mathcal{H}}{\partial \theta}
    = -2 \lambda \sqrt{I_{0}^{2}-I^{2}}\sin{\theta}
    \label{eq:I1}
\end{equation}
yields
\begin{equation}
    \frac{1}{2} \dot{I}^{2}+\mathcal{P}(I) = 0
    \label{eq:phsp}
\end{equation}
where 
\begin{equation}
\mathcal{P}(I) =
   \frac{1}{2} \left(\frac{1}{2} \chi_{0} I^{2} + (\chi 
   I_{0}+\omega ) I 
    +(\frac{1}{2} \chi_{0}I_{0}^{2}+ \omega_{0} I_{0} -E) \right)^{2} 
    - 2 \lambda^{2} (I_{0}^{2}-I^{2}) 
    \label{eq:polyn}
\end{equation}

Eq.(\ref{eq:phsp}) describes
a single nonlinear oscillator at zero energy with unit mass in a potential
$\mathcal{P}(I)$ 
\footnote{Note that we have to discard the fake time-constant solutions 
$I(t)=I_{\alpha}$, where
$I_{\alpha}$ is some zero of $\mathcal{P}(I)$, which are not solutions 
of the real dynamical system.
The reason is that the time conservation of $\mathcal{H}$ is trivial for
time-constant solutions but it is not sufficient 
to insure they fulfill the dynamical equations.}.
The solutions of eq.(\ref{eq:phsp})
oscillate in an interval  determined by two consecutive zeros of
$\mathcal{P}(I)$ and where $\mathcal{P}(I)$ is negative.
They are determined (apart from a time shift) from the
knowledge of the two invariants $E$ and $I_{0}$ and the interval
of zeros.

Since  in our case $\mathcal{P}(I)$ is a fourth degree polynomial,
the number of real zeros is even. 
Note also that all the real zeros $I_{\alpha}$ of $\mathcal{P}(I)$
($\mathcal{P}(I_{\alpha})=0$), necessarily belong to the interval
 $-I_{0} \leq I_{\alpha} \leq I_{0}$ since outside this interval
$\mathcal{P}(I)$ is the sum of two positive terms in Eq.(\ref{eq:polyn}) and 
thus cannot vanish.

The situation where $\mathcal{P}(I)$ has no zeros never occurs,
whatever the initial conditions are, because $\dot{I}^{2}$ has to be positive
at time $0$.
When $\mathcal{P}(I)$ has one pair of real zeros, there is a unique solution 
(apart from a time shift) for $I(t)$ (and $\theta(t)$) which is time-periodic. 
It corresponds to a quasiperiodic solution of the initial DNLS equation
since 
\begin{equation}
    \dot{\theta}_{0}= \frac{\partial \mathcal{H}}{\partial I_{0}}
    = -2 \lambda 
    \frac{I_{0}}{\sqrt{I_{0}^{2}-I^{2}}}\cos{\theta}
+ \chi_{0} I_{0} + \chi I + \omega_{0} .
\label{eq:theta0}
\end{equation}

When $\mathcal{P}(I)$ has two pairs of real zeros, there are two time-periodic
solutions  $I(t)$ that, again, correspond to 
quasiperiodic solutions of the initial DNLS equation.

Time-periodic solutions (corresponding to DBs) are
obtained when $I(t)$ becomes time-constant (the interval of 
oscillation shrinks to zero), i.e., when $\mathcal{P}(I)$
gets a degenerate pair of zeros.

For solutions corresponding to a targeted transfer of
energy between site $1$ and $2$, $I(t)$ should oscillate between
$+I_{0}$ and $-I_{0}$ which requires that $\pm I_{0}$
are zeros of $\mathcal{P}(I)$. This condition yields
$I_{0}=I_{T}/2$ and $E=E_{T}$ defined by Eqs.(\ref{eq:engtg}). 
Otherwise, $\mathcal{P}(I)$ must not
have any other zeros, yielding for this parameter value
\begin{equation}
\mathcal{P}(I) = \frac{1}{8}\chi_{0}^{2} 
\left(I^{2}-\frac{\omega^{2}}{\chi^{2}} 
   +\frac{16 \lambda^{2}}{\chi_{0}^{2}}\right) (I^{2} -\frac{\omega^{2}}{\chi^{2}})
  \label{eq:polyn2}
\end{equation}
which implies
\begin{equation}
    k \equiv \left|\frac{\omega \chi_{0}}{4 \lambda \chi}\right|  <1
   \label{eq:tarcond}
\end{equation}
equivalent to condition (\ref{eq:critcoup}).
Then, the solution of Eq.(\ref{eq:phsp}) for $\mathcal{P}(I)$ given by
eq.(\ref{eq:polyn2}) is a Jacobi elliptic cosine with modulus $0 \leq k <1$ 
\begin{equation}
    I(t) = -\frac{\omega}{\chi}  \mbox{cn} ( \lambda \sqrt{2} t ;k)
    \label{eq:Jacobiell}
\end{equation}
which describes explicitly the targeted energy transfer 
between the two sites. Then, the frequency of the
targeted transfer is 
\begin{equation}
    \omega_{T} =   \frac{\pi \lambda}{\sqrt{2} \mathbf{K}(k)}
    \label{eq:freqtarget}
\end{equation}
where $\mathbf{K}(k)$ is the first kind complete elliptic integral.

$ \omega_{T}$ is minimum and proportional to the coupling $\lambda$
when $\chi_{0}=0$, i.e., when the ``nonlinear tuning" between the two sites 
is optimal. Then the oscillation of the difference of action
$I(t)$ between the two oscillators becomes a pure cosine. In this case,
the explicit solution describing the optimal
targeted transfer from the donor to the acceptor obtained for 
$E=E_{T}=-\omega \omega_{0}/\chi$ and $I_{0}=I_{T}/2=-\omega/\chi$ is rather simple

\begin{eqnarray}
    \psi_{1}(t) & =  & \sqrt{\frac{-\omega}{\chi}} \cos{\lambda t}
    \exp{-i(\frac{1}{4\lambda} \sin{2\lambda t}+\frac{\omega_{0}t}{2})} 
    \label{eq:psi1}  \\
    \psi_{2}(t) & = & i \sqrt{\frac{-\omega}{\chi}} \sin{\lambda t}
    \exp{-i(\frac{1}{4\lambda} \sin{2\lambda t}+\frac{\omega_{0}t}{2})} 
    \label{eq:psi2}
\end{eqnarray}

\begin{figure}[tbp]
    \centering
 \includegraphics{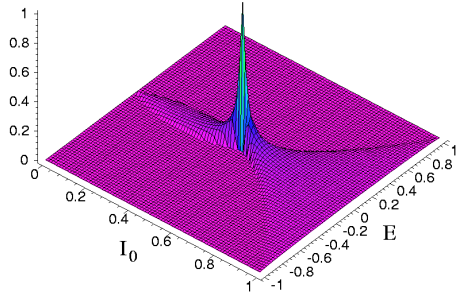}
    \caption{3D plot of the rate $A$ of transfer for the action 
    between the two sites of the dimer versus $E$ and $I_{0}$ for  
   Hamiltonian (\ref{eq:energ}) at $\chi_{0}=0$,$\chi=2$, 
   $\omega_{0}=0$, $\omega=-1$. Coupling $\lambda=0.01$ is weak.
   $100$\% transfer occurs at the top of the peak
   while away from the peak, action transfer becomes negligible.}
    \label{fig2}
\end{figure}

Fig.\ref{fig2} shows the 3D plot of the rate of action transfer
$A=(\max_{t}I(t) -\min_{t}I(t))/(2I_{0})$ 
between the two sites as a function of the energy $E$ and the 
total action $I_{0}$ of the solution.

The targeted energy transfer solution disappears when its frequency 
reaches zero, i.e., $k$ becomes equal to $1$. Then, polynomial 
$\mathcal{P}(I)$ gets a degenerate zero at $I=0$ which means that a 
time-periodic solution at $I=0$  appears. It corresponds to a new DB 
solution which suddenly blocks the targeted transfer.
At the limit $k=1$, the elliptic cosine becomes a sech function 
\cite{KC87}
and close to this limit, the transfer of energy becomes
intermittent.

The time-periodic solutions (DBs and multibreathers)
are characterized by  $\dot{I}=0$ in Eq.(\ref{eq:I1})
 and $\dot{\theta}= \partial \mathcal{H}/\partial I =0$ 
yields $\theta=0$ or $\pi$ and 
 \begin{equation}
     4 \lambda^{2} I_{b}^{2} - (\chi_{0} I_{b} 
     + \chi I_{0} +\omega)^{2}(I_{0}^{2}-I_{b}^{2}) =0 .
     \label{eq:brsol}
 \end{equation}
This is a fourth degree equation which may have $2$ or $4$ real zeros,
functions of $I_{0}$. The energy of these solutions $E_{b}(I_{0)}$
determines the boundaries of the domain in the plane of initial conditions
$E,I_{0}$ that can be realized.

\begin{figure}[tbp]
    \centering
 \includegraphics[angle=270]{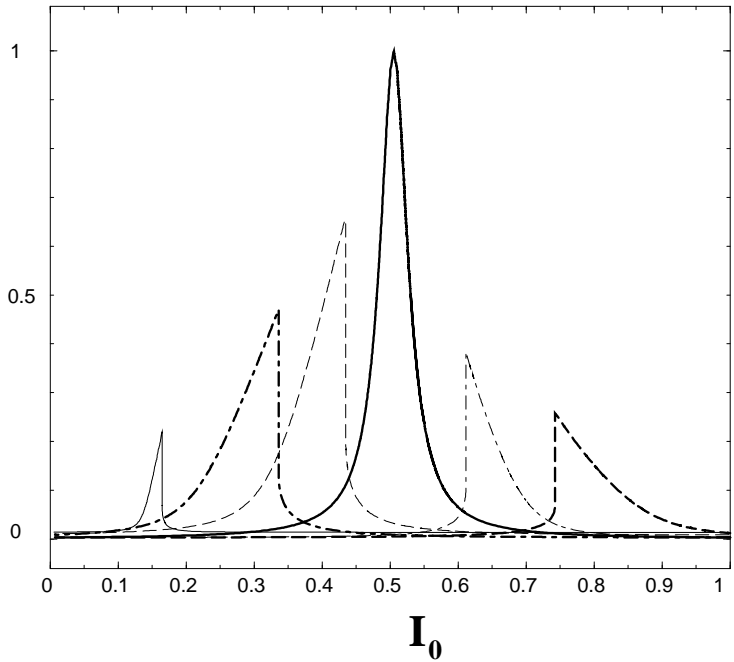}
    \caption{Transfer rate in action versus $I_{0}$ for
    $\chi=2$, $\omega=-1$, $\lambda=0.01$ and several values of
    $\chi_{0}=$ $-0.8$ (thick dashed), $-0.5$ (thin dashed-dotted),
$0$ (thick full line),  $0.25$ (thick dashed),
    $0.5$ (thin dashed-dotted), $5.$ (thin full).}
    \label{fig3}
\end{figure}

We can also analyze partial energy transfer by considering
the set of initial conditions where all the energy is at site $1$.
These conditions are characterized by
$I(0)=I_{0}$ and  $E=(\chi_{0}+\chi)I_{0}^{2}+(\omega+\omega_{0})I_{0}$.
Then, polynomial (\ref{eq:polyn}) becomes
\begin{equation}
\mathcal{P}(I) = \frac{1}{2} (I-I_{0})\left( (I-I_{0})
(\frac{1}{2} \chi_{0} (I+I_{0}) + \chi I_{0}+\omega)^{2} 
    + 4 \lambda^{2} (I_{0}+I)\right) .
    \label{eq:polyn3}
\end{equation}
Then, $I(t)$ oscillates between $I_{0}$ and the largest zero $I_{m}$
of this polynomial. Fig.\ref{fig3} shows the rate of transfer in action
$(I_{0}-I_{m})/(2I_{0})$ versus $I_{0}$ for a given set of parameters 
$\chi$,$\omega$, $\lambda$ and for several values of $\chi_{0}$.
100\% transfer is obtained at $I_{0}=I_{T}/2=1/2$ when condition
(\ref{eq:critcoup}) is fulfilled, i.e., $|\chi_{0}|<4 |\lambda 
\chi|/|\omega|= 0.08 $. There are discontinuities on these curves 
where the oscillation regime of $I(t)$ becomes cnoidal which
are due to the appearance of new time-periodic solutions 
blocking the energy transfer.

\section{Detecting targeted energy transfer in complex models}
This theory of targeted transfer can be used in practice in complex 
systems to detect targeted transfer because the functions involved
in the effective dimer Hamiltonian (\ref{eq:hameff}) can be
explicitly calculated. For that purpose, the initial system has to be 
separated into two coupled subsystems.
The first one will sustain the donor DB and the second one 
the acceptor DB.
The coupling between the two systems is not necessarily
weak if the considered DBs are far from each other
but the effective coupling has to be weak.

Using the numerical techniques  developed earlier for the
calculations of DBs in arbitrarily complex systems \cite{KA00,KAI99,KAII00}, 
one chooses a DB solution in the isolated donor system
and calculate its dynamical configuration
$\{p_{i}^{D}(\omega_{1}t),q_{i}^{D}(\omega_{1}t)\}$,
its energy, and its action. Then, 
varying the frequency, one obtains function $H_{D}(I)$.
The same can be done for a DB in the acceptor system.
For a complete targeted transfer, we require that both DBs be 
gapless, i.e., $H_{D}(I)$ and $H_{A}(I)$ can be continued to $I=0$ where 
$H_{D}(0)=H_{A}(0)=0$.

Since the \textit{effective} coupling between the two DBs is 
supposed to be small,
$I_{T}$ can be determined approximately by the equality
$H_{D}(I_{T})=H_{A}(I_{T})$,
which requires that the two curves intersect at $I\neq 0$.
For having targeted transfer at reasonably small coupling,
function  $\epsilon(I)=H_{D}(I_{T}/2+I)+H_{A}(I_{T}/2-I)$
should be small. Once these conditions are fulfilled,
we have to check that our dimer assumption is valid, i.e.,
that there is no other DB either on the donor system
or the acceptor system with almost the energy functions $H_{D}(I)$
or $H_{A}(I)$ in the interval $[0,I_{T}]$.

Next, the coupling energy between the two systems
is considered. It is originally a function of the global system configuration
$\{p_{i}^{D},q_{i}^{D}\}$ and  $\{p_{i}^{A},q_{i}^{A}\}$
It yields the effective coupling in the dimer model by averaging
this energy over 
the fast variable $\theta_{0}$ for the system configuration
$\{p_{i}^{D}((\theta_{0}+\theta)/2),q_{i}^{D}((\theta_{0}+\theta)/2)\}$
and $\{p_{i}^{A}((\theta_{0}-\theta)/2),q_{i}^{A}((\theta_{0}-\theta)/2)\}$.
It is thus only a function of $I_{0}$, $I$ and $\theta$
which has to be small.

Comparing the precise detuning function and
the coupling functions as in Eq.(\ref{eq:detuning}),
we can predict rather accurately the possible existence  of targeted 
transfer solutions between these  donor and acceptor DBs.
In the cases we tested up to now, where these conditions are 
correctly fulfilled, molecular dynamics confirmed 
the existence of targeted transfer where almost 100\% of the energy 
oscillates for a very long time between the donor and the acceptor DB.
As expected also, this energy transfer is blocked as soon as the initial 
amount of energy is not optimal or the coupling too weak \cite{KAT00}.

\section{Irreversibility in the energy targeted transfer}

DBs are out of equilibrium excitations. In real systems, which are
not strictly at 0 K, they interact with the phonon modes and other
excitations which should make their lifetime finite.
Thermalization should be in principle a consequence of intrinsic 
nonlinearities of the system. However, standard approaches usually
simulate thermalization and relaxation  by a white Langevin  noise
and an extra damping. This is known to be equivalent to a coupling 
of the initial system with a fictitious extra system consisting of 
an infinite set of harmonic oscillators with a uniform distribution 
of frequencies \cite{Zwa73}. This is clearly a bias which makes the DBs 
to systematically radiate energy in this bath and 
artificially washes out their existence as very long lifetime solutions.
Actually, the long lifetime of DBs is precisely 
due to the fact that they have no resonance with the linear phonon spectrum
which implies that either their frequencies are above the phonon 
spectrum or there are gaps in this phonon spectrum. It is the nonuniform
and specific phonon spectrum of the model itself which should be
considered for studying DB lifetime as we started to do in some
recent works \cite{JA00,RABT00}. In other words, real systems exhibit a 
selective damping depending on the frequencies of their modes
which should be explicitly taken into account to be physically 
correct. Thus, the thermal bath assumption with a uniform phonon spectrum 
is not physically founded and, worst, induces the denaturation of the 
model.

We discussed above the ideal case for targeted energy transfer
at 0 K. For obtaining an ideal transfer, besides the adiabaticity hypothesis, 
the donor and acceptor DBs must be gapless.
This condition requires that there exist no linear modes in the 
donor (or the acceptor) system with frequency in the range  of  variation
of the DBs frequency. Indeed, it has been shown in \cite{KA00,KAI99,KAII00} that
before the frequency
of a DB collides with a linear mode frequency, it systematically 
bifurcates with another one. In this case, the DB  is not strictly
continuable to zero amplitude. There is a frequency 
discontinuity corresponding to the region where the resonance with 
the linear mode destroys the DB. 
Strictly speaking, when there is a linear mode in the range of 
variation of frequency of a DB on the donor (or acceptor) system,
this DB exhibits a nonvanishing energy gap which in principle invalidates
our assumptions for describing the coupled system as a generalized 
dimer model (\ref{eq:hameff}) .
As a consequence, the frequency of the ideal targeted transfer solution
precisely oscillates in between the edges of a gap of the linear 
phonon spectrum.

We would like to briefly discuss here what can happen to targeted energy 
transfer if there is just a single resonant linear mode in 
the phonon gap, first in the case of strong coupling, i.e., this linear 
mode is spatially located close to the donor (or the acceptor) DBs.
As we just said, the DB of the donor (or the acceptor) will exhibit
an energy gap, so that the detuning function cannot be defined
for the whole interval of variation of $I$.
The energy transfer will become incomplete and irreversible.
As the frequency of the targeted energy
transfer solution approaches the linear resonance, a substantial fraction of the
energy will be suddenly transferred to this linear mode instead of the 
acceptor DB. Consequently, the 
frequency of the donor DB will change independently of the 
acceptor DB. The loss of resonance between donor and acceptor DB stops
immediately the energy transfer. If one assumes that the
energy captured by the linear resonant mode does not return 
significantly to the  donor DB, the part of the energy left on the
donor system and of the linear mode, could be 
dissipated into phonons through chaotic processes and 
finally heat. Then, only  a well-defined fraction
of the initial energy on the donor has been transferred and the excess
of energy has been dissipated into heat. We obtained only an 
incomplete but irreversible targeted  transfer.

The effect should be much less sharp when the linear mode which is resonant 
with the donor DB is weaker, i.e., the resonant linear mode is physically located
far away from the  donor and acceptor DBs, say, in order to fix the ideas,
on the donor system. Then, there is only a small discontinuity gap in 
frequency to jump for the donor DBs where function $H_{D}(I_{D})$ is 
undefined. We think that despite a small energy loss when 
crossing the resonance,  targeted energy transfer could be continued
beyond this resonance and be almost completed. If there are many such 
weak resonances they could be described as a frequency dependent 
damping depending on the distribution and coupling with resonant
linear modes. Resonances with bands of extended 
phonons should have a similar effect.

Actually, in realistic systems there are likely some linear modes
in the phonon gap causing  some energy dissipation
and making energy transfer irreversible. On the opposite,  at finite
temperature, thermal activation of the linear modes could also help
targeted energy transfer by stochastic resonance effects. 
These problems will be investigated in the future.

\section{Concluding remarks and new perspectives}
We have demonstrated here that the essential difference between linear
and nonlinear resonance (allowing the targeted energy transfer) is that nonlinear
resonance is \textit{selective} in energy (or action), unlike linear
resonance, which is not.  Targeted energy transfer occurs  because
the frequencies of the donor oscillator or DB and acceptor oscillator or DB,
although they both vary during the transfer, remain almost equal.
This condition can be achieved at weak coupling, only for a certain
initial energy  $E_{T}$  and when  a certain detuning function is
small enough compared to the coupling energy.  
In the ideal case,  an amount of energy $E_{T}$
deposited on the donor DB oscillates slowly and periodically back and forth
between donor and acceptor. It is totally transferred between the donor
and the acceptor DB at periodic times.
Otherwise, although there is some energy oscillation,
the energy transfer becomes negligible.

We do not believe that the obtained targeted energy transfer 
solutions correspond to strictly exact solutions of the
initial model because of the hypothesis of adiabaticity
which cannot be perfectly fulfilled.
However, situations close to the ideal case could exist
where the number of energy oscillations between donor and acceptor DBs
is very large. The high sensitivity to the initial energy
makes that this energy transfer can be easily stopped by intermediate 
linear resonance which can be viewed as a kind of transistor effect.
This effect could be used in practical devices (e.g., with 
optical fibers).

Within the perspectives opened up by this approach, we would like to 
suggest a series of new physical problems to be investigated in the near future.

We may consider situations where, instead of one, there are several DBs
on the donor system with almost identical functions $H_{D}(I)$ and
 one acceptor DB on the acceptor system . Then, if we are in the 
 condition of targeted energy  transfer for a single donor DB,  
any  donor DBs excited at the energy $E_{T}$ will transfer their energy
to the acceptor DB. However, the transfer has good chances to be 
irreversible because when the transfer is completed, the acceptor energy could 
return simultaneously to all donor DBs but then the resonance between
the donor DBs and the acceptor DB will be broken and the transfer will be 
blocked. This situation is interesting for realizing energy funneling
analogous to what is observed in chlorophyll \cite{HDRS98}.

We may also consider situations such that for an energy 
$E_{T}^{\prime}<E_{T}$, the acceptor DB (A) is tuned to another acceptor DB
($A^{\prime}$). Then, before the transfer from the initial donor (D)
to the acceptor (A) is completed, this acceptor (A) transfer its 
energy to the next acceptor ($A^{\prime}$) with some energy loss
(note that the DB $A^{\prime}$ acts initially like an intragap linear 
resonance) but instead of dissipating energy, it may collect most of 
the $A$ acceptor energy.
By this way, recurrent transfer could be built along a certain
well determined sequence of acceptor DBs. We could then realize irreversible
cascades of energy transfer (with some energy losses however).
It is also important to realize now that 
DBs in molecules may have piezoactive effects, i.e., they might generate
deformation of the molecules as emphasized in \cite{Aub98}. 

This effect could be important for example for understanding the
energy conversion of ATP into mechanical energy (molecular motors).
The energy deposition from ATP could occur first as a DB located at a
certain receptive site of the moving biomolecule (e.g., kinesin adsorbed on a
microtubule). Then, this DB could cascade by targeted energy transfer through
a series of DBs located at specific sites of the biomolecule generating
a well determined sequence of deformations which lead to a well determined 
motion of the biomolecule (step) along the oriented microtubule
\footnote{Microtubules are polar biopolymers. 
They often give rise to ratchet models.} after the time $t_{m}$ 
necessary for this cascade process. Beyond this time,
the DB energy is finally relaxed into heat. 
Thus, excitations by ATP randomly repeated at  time intervals
longer than this characteristic time 
$t_{m}$ should produce systematically a directive walk of this biomolecule along 
the microtubule (this motion should become randomly intermittent in time
if the ATP concentration is low). 
In the future, we plan to use and to develop our tools for calculating 
DBs in  nonlinear systems for controlling precisely their targeted energy
transfer, funneling, cascades, piezoactivity of DBs and thus for
designing artificial models with desired properties mimicking those of
biosystems.

Let us also note that instead of being atomic vibrations, these DBs could be 
electronic vibrations (excitons), polarons, or atomic vibrations combined with 
an electronic charge (polarobreathers) \cite{Aub97}. Thus, there are many 
possible variations of the targeted transfer principle depending on the 
considered physical problem (transfer of energy,  charge, spin, 
etc) and its scale of energy. This theory of targeted
transfer could be useful, for example, for revisiting the 
standard theories of fluorescence (\cite{Lak83}), chemical reactions, catalysis,etc.

\begin{ack}
S.Aubry would like to dedicate his contribution to this work
to the memory of his mother who recently disappeared.

  We thank R.S. MacKay for frequent and fruitful discussions 
  about dynamical systems. We also acknowledge  M. Johansson
  for his interest and comments concerning this work.
  G. Kopidakis acknowledges support by Greek G.S.R.T. (E$\Pi$ET II).
  This work has been supported by EC under contract HPRN-CT-1999-00163.

\end{ack}


\begin{thebibliography}{99}
\bibitem{And58} P.W. Anderson, Phys. Rev.~\textbf{109} (1958) 1492.
\bibitem{Economou}C. M. Soukoulis, E. N. Economou,
     Waves Random Media \textbf{9} (1999) 255.
\bibitem{KA00} G. Kopidakis, S. Aubry, Phys. Rev. Lett. \textbf{84} (2000) 
     3236.
\bibitem{KAI99} G. Kopidakis, S. Aubry, Physica \textbf{D 130} (1999) 
     155. 
\bibitem{KAII00} G. Kopidakis, S. Aubry, Physica \textbf{D 139} (2000) 247.
\bibitem{ST88}
A. J. Sievers, S. Takeno (1988) Phys. Rev. Lett. \textbf{61} 970.
\bibitem{CP90} D.K. Campbell, M. Peyrard in: D.K. Campbell (Ed.),
\textit{CHAOS-Soviet American Perspectives on Nonlinear Science},
American Institute of Physics, 1990.
\bibitem{Lan33} L. Landau, Phys. Z. Sowjetunion \textbf{3} (1933) 664.
\bibitem{BS26} R.T. Birge, H. Sponer, Phys. Rev. \textbf{28} (1926) 
259. 
\bibitem{E29} J.W. Ellis, Phys. Rev. \textbf{33} (1929) 27.
\bibitem{OE82} A.A. Ovchinnikov, N.S. Erikhman, Usp.Fiz.Nauk
      \textbf{138} (1982) 289; Sov. Phys. Usp. \textbf{25} (1982) 738.
\bibitem{Far96} S.C. Farantos, Int. Rev. Phys. Chem. \textbf{15} (1996) 
345.
\bibitem{Sco99} Alwyn Scott \textit{Nonlinear Science, Emergence and Dynamics of
Coherent Structures}, Oxford University Press, Oxford, 1999.
\bibitem{FW98} S. Flach, C.R. Willis, Phys. Rep. \textbf{295} 
(1998) 182.
\bibitem{MA94} R.S. MacKay, S.Aubry, Nonlinearity \textbf{7} (1994) 1623.
\bibitem{MS95} R.S. MacKay, J-A. Sepulchre,
Physica \textbf{D 82} (1995) 243.
\bibitem{Aub97}S. Aubry, Physica \textbf{D 103}, (1997) 201-250.  
\bibitem{SM97} J-A. Sepulchre, R.S. MacKay, Nonlinearity 
\textbf{10} (1997) 679.
\bibitem{LSM97} R. Livi, M. Spicci, R.S. MacKay, Nonlinearity 
\textbf{10} (1997) 1421. 
\bibitem{Aub98} S. Aubry, Ann. Inst. H. Poincar\'e, Phys. Th\'eor.
\textbf{68} (1998) 381.
\bibitem{MA96} J.L. Mari\~n, S. Aubry, Nonlinearity 
\textbf{9} (1996) 1501.
 \bibitem{CAF98} T. Cretegny, S. Aubry, S. Flach,
Physica \textbf{D 119} (1998) 73.
\bibitem{JA00} M. Johansson, S. Aubry,
 Phys.Rev. \textbf{E 61} (2000) 5864.
 \bibitem{CAT96} Ding Chen, S. Aubry, G. Tsironis,
Phys. Rev. Letts \textbf{77} (1996) 4776.
\bibitem{AC98} S. Aubry, T. Cretegny,
Physica  \textbf{D 119} (1998) 34.
 \bibitem{TA96} G.P. Tsironis, S. Aubry, Phys. Rev. Lett. \textbf{77} (1996) 5225.
\bibitem{BVAT99} A. Bikaki, N.K. Voulgarakis, S. Aubry, G.P. 
Tsironis, Phys. Rev. \textbf{E 59} (1999) 1234.
\bibitem{RABT00}K.\O. Rasmussen, S. Aubry, A.R. Bishop, G.P. 
Tsironis, Eur. Phys. J. \textbf{B 15} (2000) 169.
  \bibitem{ESMBA98} H.S. Eisenberg, Y. Silberberg, R. Morandotti, 
 A.R. Boyd, J.S. Aitchison, Phys. Rev. Lett. \textbf{81} (1998) 3383.
 \bibitem{MPAES99} R. Morandotti, U. Peschel, J.S. Aitchison,
 H.S. Eisenberg, Y. Silberberg,  Phys. Rev. Lett. \textbf{83} (1999) 
 2726.
 \bibitem{TMO00} E. Trias, J.J. Mazo, T.P. Orlando, Phys. Rev. 
 Lett. \textbf{84} (2000) 741.
 \bibitem{BAUFZ00} P. Binder, D. Abraimov, A.V. Ustinov, S.Flach,
 Y. Zolotayuk, Phys. Rev. Lett. \textbf{84} (2000) 745.
\bibitem{SBLSS99} B.I. Swanson, J.A. Brozik, S.P. Love, G.F. Strouse,
A.P. Shreeve, A.R. Bishop, W-Z. Wang, M.I. Salkola,
Phys. Rev. Lett. \textbf{82} (1999) 3288.
\bibitem{SES99} U.T. Schwartz, L.Q. English, A.J. Sievers,
Phys. Rev. Lett. \textbf{83} (1999) 223.
\bibitem{XMA00} A. Xie, L. van der Meer, W. Hoff, R.H. Austin, 
Phys. Rev. Lett. \textbf{84} (2000) 5435.
\bibitem{Aub98b} S. Aubry, \textit{Discrete Breathers in Nonlinear Lattices:
New Perspectives in Physics and Biological Physics}, (review) (1998) unpublished.
\bibitem{AMS00} T. Ahn, R.S. MacKay, J-A. Sepulchre 
\textit{Dynamics of relative phases: generalised multibreathers},
Nonlinear Dynamics, to appear.
\bibitem{FKM97} S. Flach, K. Kladko, R.S. MacKay, 
Phys. Rev. Lett. \textbf{78} (1997) 1207.
\bibitem{FAM98} J.L. Floria, J.L. Mari\~n, S. Aubry, P.J. 
Martinez, F. Falo, J.J. Mazo, Physica \textbf{D 113} (1998) 387.
\bibitem{KC87} V.M. Kenkre, D.K. Campbell, Phys. Rev.\textbf{B 34} 
(1987) 4959.
\bibitem{KT87} V.M. Kenkre, G.P. Tsironis, Phys. Rev.\textbf{B 35} (1987) 1473.
\bibitem{GS98} B.C. Gupta, P.A. Sreeram, Phys. Rev. \textbf{B 57} (1998) 4358.
\bibitem{KAT00} G. Kopidakis, S. Aubry, G. Tsironis,
submitted to Phys. Rev. Lett.
\bibitem{Zwa73} R. Zwanzig, J.Stat.Phys. \textbf{9} (1973) 215.
\bibitem{HDRS98} X. Hu, A. Damjanovic, T. Ritz, K. Schulten, 
P. Natl. Acad. Sci. (USA) \textbf{95} (1998) 5935.
\bibitem{Lak83} J. Lakowicz, \textit{Principles of Fluorescence
 Spectroscopy}, Plenum Press, New York, 1983.

\end{thebibliography}
\end{document}